\documentclass[12pt]{article}
\pdfoutput=1
\usepackage[numbers]{natbib}
\usepackage{amsmath}
\usepackage{graphicx}
\usepackage{enumerate}
\usepackage{xcolor}
\usepackage{booktabs}
\usepackage{adjustbox}
\usepackage{url} 
\usepackage{hyperref}

\newcommand{\blind}{1}

\renewcommand{\arraystretch}{0.8}

\begin{document}

\def\spacingset#1{\renewcommand{\baselinestretch}%
{#1}\small\normalsize} \spacingset{1}


\if1\blind
{
  \title{\bf Functional Clustering of Neuronal Signals with FMM Mixture Models}
  \author{Alejandro Rodríguez-Collado\thanks{The authors gratefully acknowledge the financial support received by the Spanish Ministerio de Ciencia e Innovación [PID2019-106363RB-I00 to AR-C and CR]. This study was also partially supported by Fundación Eugenio Rodríguez Pascual (Biomedical Research Grants 2021).}
  \thanks{Corresponding Author: alejandrorodriguezcollado@gmail.com}\hspace{.2cm}\\
    Department of Statistics and Operations Research, Universidad de Valladolid\\
    and \\
    Cristina Rueda$^*$ \\
    Department of Statistics and Operations Research, Universidad de Valladolid}
  \maketitle
} \fi

\if0\blind
{
  \bigskip
  \bigskip
  \bigskip
  \begin{center}
    {\LARGE\bf Functional Clustering of Neuronal Signals with FMM Mixture Models}
  \end{center}
  \medskip
} \fi






\bigskip
\begin{abstract}
The identification of unlabelled neuronal electric signals is one of the most challenging open problems in neuroscience, widely known as Spike Sorting. Motivated to solve this problem, we propose a model-based approach within the mixture modeling framework for clustering oscillatory functional data called MixFMM. The core of the approach is the FMM waves, which are non-linear parametric time functions, flexible enough to describe different oscillatory patterns and simple enough to be estimated efficiently. In particular, specific model parameters describe the waveforms' phase, amplitude, and shape. A mixture model is defined using FMM waves as basic functions and gaussian errors, and an EM algorithm is proposed for estimating the parameters. In addition, the approach includes a method for the number of clusters selection. Spike Sorting has received considerable attention in the literature, and different functional clustering approaches have traditionally been considered. We compare those approaches with the MixFMM in a broad collection of datasets, including benchmarking simulated and real data. The MixFMM approach achieves outstanding results in a selection of indexes across datasets, and the significant improvements attained in specific scenarios motivate interesting neuronal insights.
\end{abstract}

\noindent%
\vfill

\newpage
\spacingset{1.9} 
\section{Introduction}
\label{sec:intro}
The analysis of the electric activity of the neurons is regarded as one of the most practical and effective approaches to studying the nervous system. The electric signals recorded by electrodes register rapid voltage rises lasting a few milliseconds called spikes. The voltage returns to the initial baseline level in each spike, describing a single oscillation. The study of spikes is crucial in neuroscience, as spikes serve as the informational unit between neurons, and their firing rate and shape determine the cell’s morphological, functional, and genetic type.  In particular, spikes fired by a particular neuron recorded under similar conditions, such as the electrode’s distance and orientation, are assumed to have a specific shape. Spike Sorting (SS) is the collection of techniques to identify spikes corresponding to different neurons. The correct identification of spikes is crucial for studying the connectivity patterns between close-by neurons \cite{Buz04}, relating the firing of certain neurons to the memory process \cite{Rey15}, the treatment of epileptic patients \cite{Sha17}, or the development of high-accuracy brain-machine interfaces \cite{Mog19}, among many other questions. However, SS remains one of the most challenging open problems in neuroscience. The main reasons are the low signal-to-noise ratio, the waveform variability of a particular neuron's spikes, the selection of adequate features to characterize the spike shape, or the overlapping of spikes fired nearby in time. Algorithms that simultaneously address all of these issues are an ever-increasing necessity, as manual spike curation has become unaffordable in light of the increased data availability in the next generation of recording techniques \cite{Rey15}.

\begin{figure}[!ht]
\begin{center}
\includegraphics[width=0.9\textwidth]{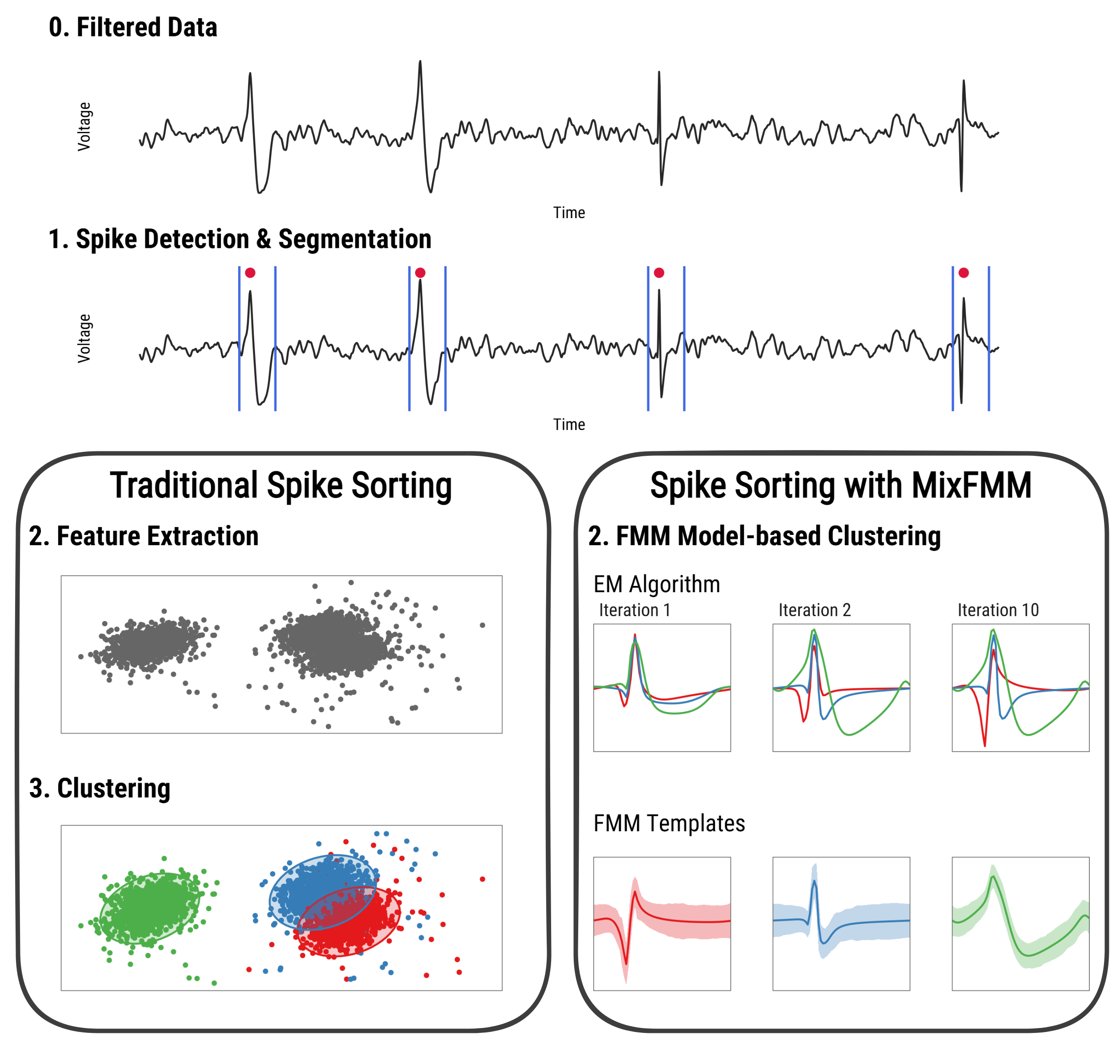}
\end{center}
\vspace{-10mm}
\caption{Overview of SS: stages 0 and 1 (top), stages 2 and 3 in traditional SS procedures (bottom left), and single model-based clustering stage in MixFMM model (bottom right).\label{fig:Figure1}\vspace{-1mm}}
\end{figure}

The four traditional stages of SS are shown in Figure \ref{fig:Figure1}, top and bottom left. In stage 0, voltage traces are preprocessed, often with a basic bandpass filtering. Stage 1 entails the automatic spike detection and the signal segmentation into various sub-signals, each containing a single spike. Finally, stages 2 and 3 are devoted to the extraction of discriminant features and clustering, respectively. The focus of this paper is on stages 2 and 3, which may be quite different from one approach to another; features are mainly extracted using principal component analysis (PCA) or  wavelet coefficients, while the most widely used clustering algorithms are k-means (KM) and  Gaussian mixture models (GM). However, a complete list of alternative techniques used in stages 2 and 3 in SS is large. Some interesting references are \cite{Qui04, Eka14, Rey15, Car18, Sou19, Rac19, Car19, Suk19, Wan19, Vee20}.

We propose in this paper an original parametric model-based functional clustering approach as an alternative to stages 2 and 3 in traditional SS (Figure \ref{fig:Figure1}, bottom right).
  
Model-based functional clustering is an active topic in the literature \citep{Jac14, Giaf13, Jac14, Lee19, Lim19, Zho20}. Different parametric models have been studied, mainly regression models and additive models defined by basic time functions such as splines, Fourier waves, or wavelets. The model choice depends on the available information, in terms of explicative variables, on the nature of the data, and on the sources of variation characterizing data, among other issues. For instance, Fourier methods might not be suitable for nonstationary signals. Specifically, two main sources of variation are considered in functional analysis, the \textit{amplitude variation} that measures vertical variability, and  the \textit{phase variation} measuring horizontal (lateral) displacements \cite{Mar15,Juh17, Cla21}. Depending on the problem at hand, either both sources of variation are the focus or one is just nuisance and must be removed. Besides, some methods do not work adequately without prior correction for phase differences. \vspace{-1mm}\\

The model-based approach suitable for analyzing oscillatory signals that we present is the FMM mixture (MixFMM). The core is a mixture of gaussian distributions where the means are sums of FMM waves or components. FMM waves are non-linear time functions that describe a single oscillation with four parameters $(A,\alpha,\omega,\beta)$, which characterize the amplitude ($A$), phase ($\alpha$), and shape ($\beta, \omega$) of the waveform. In particular, the amplitude and phase source of variation issues can be easily handled using this model. Moreover, the mean cluster waveforms or templates can be compared in a simple way. Signals defined as combination of FMM waves have been successfully used to analyze oscillatory signals arising in different fields, such as ECG and gene expression data, as well as neuronal spikes \cite{Rue19,Rue21b,Rod21a,Rod21b}. In all of these works, the method attains highly accurate predictions and allows the extraction of interpretable features, among other assets.  The number of components is often associated with the prominent peaks and troughs in the signal, these being five for ECG signals or three for neuronal spikes. The first component, also called the dominant component, usually identifies the prominent underlying biological process and, in some contexts, explains most of the data variability, as in neuronal spikes. Finally, another remarkable property of FMM waves, very useful in functional analysis, is that they are truly dynamic that can be formulated as an ordinary differential equations system.

Noise existing in data is taken into account using FMM$_m$ models, defined as signal plus error models where the signal is a combination of $m$ FMM waves and the error is assumed gaussian. The maximum likelihood estimator (MLE) of the parameters are derived using a backfitting algorithm, in which FMM$_1$ models are repeatedly fitted to the residue until a stop criterion is attained \cite{Rue21a}. The novel approach is based on MixFMM$_m$ models defined as an FMM$_m$ mixture. We solve the MixFMM$_m$ model's parameter estimation with an expectation-maximization (EM) algorithm designed ad-hoc that makes use of the FMM$_m$ estimation algorithm. In addition, we propose a likelihood-based method to select the number of clusters. The MixFMM approach has been validated in 10 benchmarking SS datasets (8 simulated and 2 real), including a wide variety of spike waveforms, noise levels, and recording situations. The results are compared with two standard approaches in SS, PCA plus either KM or GM, using indexes based on the associated groundtruth and internal cluster metrics. The MixFMM achieves outstanding results, clearly superior to its competitors in terms of accuracy, interpretability, robustness, cluster cohesion and separability. Moreover, the proposed method estimates the number of clusters better than other traditional approaches. \vspace{-1mm}\\

The rest of the paper is organized as follows. Section \ref{sec:Methods} introduces the MixFMM$_m$ model, the estimation algorithm, and the number of clusters selection. Section \ref{sec:ssDatasets} describes the datasets analyzed, and Section \ref{sec:NumStudies} presents the main results. Finally, in Section \ref{sec:Discussion}, concluding remarks and future work are discussed.

\section{Methods}
\label{sec:Methods}
\vspace{-3mm}
In the following, let us assume that $\boldsymbol{t}=(t_1, ... ,t_p)$, with $t_0 \le t_{1}<t_{2}<...<t_{p} \le T$, and $\boldsymbol{x}_1,...,\boldsymbol{x}_n$, $\boldsymbol{x}_i=(x_i(t_1),...,x_i(t_p)), \forall i \in \{1,...,n\}$ are the time points and observed signals, also called curves. Without loss of generality, we assume that $t\in [0,2\pi)$. Otherwise, consider $t'=\frac{(t-t_0)2\pi}{T}$. 
\vspace{-4mm}
\subsection{FMM background}
\begin{figure}[!ht]
\begin{center}
\includegraphics[width=1\textwidth]{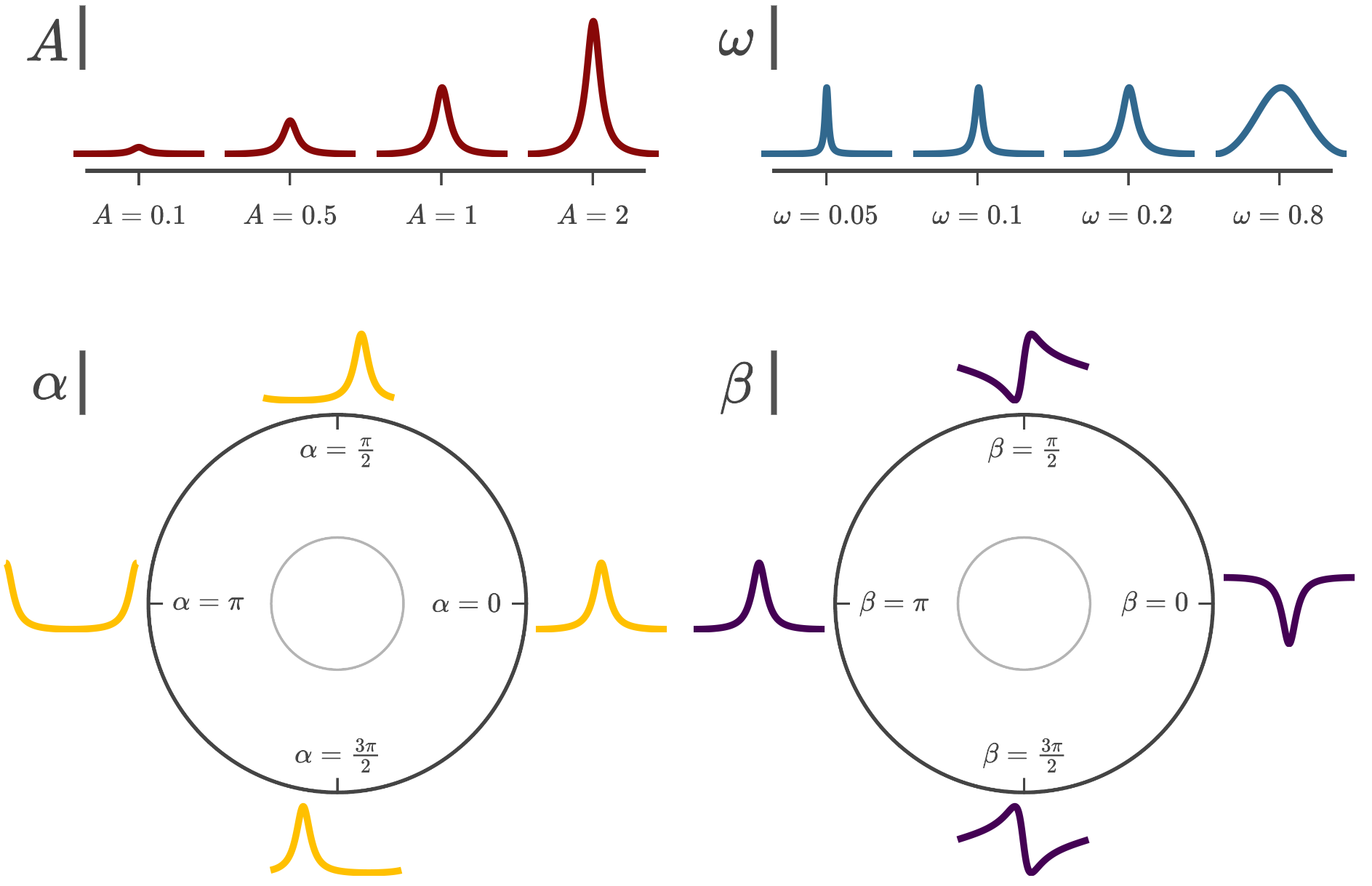}
\end{center}
\vspace{-8mm}
\caption{$W(t;A,\alpha,\beta,\omega)$ for various $(A,\alpha,\beta,\omega)$. Unless stated otherwise, $A=1, \alpha=0, \beta=\pi, \omega=0.2$. \label{fig:Figure2}}
\end{figure}

An FMM wave is defined as follows,
\vspace{-3mm}
\begin{equation}
	W(t;A,\alpha,\beta,\omega)=A\cos\left(\beta+2\arctan\left(\omega\tan\left(\frac{t-\alpha}{2}\right)\right)\right).
\end{equation}
\vspace{-2mm}
Specifically,  $A \in \Re^+$ and $\alpha \in [0,2\pi)$ are measures of  the wave amplitude and phase location, respectively, while $\beta \in [0,2\pi)$ and $ \omega \in [0,1]$ describe the shape. Figure \ref{fig:Figure2}  shows the waveform patterns for different parameter configurations.

The multicomponent FMM model of order $m$, FMM$_m$, is a Gaussian model in which the mean, $\mu(t;\boldsymbol{\theta})$, is defined as a sum of FMM waves as follows:
\begin{equation}\label{ec:1}
 \mu(t;\boldsymbol{\theta})= M+ \sum_{J=1}^m W(t;A_J,\alpha_J,\beta_J,\omega_J)
\end{equation}
where $\boldsymbol{\theta}=(M,A_1,\alpha_1,\beta_1,\omega_1,...,A_m,\alpha_m,\beta_m,\omega_m)$ verifies:
\begin{enumerate}
		\item $ M \in \Re $; $(A_J,\alpha_J,\beta_J,\omega_J) \in \Theta_J= \Re^+ \times [0,2\pi) \times [0,2\pi) \times[0,1]$; $J =1,...,m$
		\item $\alpha_{1} \le \alpha_2 \le ... \le \alpha_{m-1} \le \alpha_{m} \le \alpha_{1}$.
		\item $A_{1} = \max_{1 \leq J \leq m} A_J.$
\end{enumerate}

The restrictions guarantee the identifiability of the model parameters and provide biophysiologically interpretable solutions. For instance, the $\alpha$ restrictions correspond to the assumption that the atrial depolarization is previous to the ventricle depolarization in the analysis of ECG signals, and that the cell membrane repolarization precedes its hyperpolarization in neuronal spikes \cite{Rue19, Rue21a}.

\subsection{FMM Mixture Models}
The MixFMM$_{m}$ is defined as a mixture model with density defined as follows:
\begin{equation} \label{eq:mixFmmDensity}
f(\boldsymbol{x}|\boldsymbol{\Psi})=\sum_{k=1}^K \gamma_k \; \mathcal{N}(\boldsymbol{x}; \boldsymbol{\mu}(\boldsymbol{t};\boldsymbol{\theta}_k), \sigma^2_k \boldsymbol{I}_p),
\end{equation}
where $\boldsymbol{I}_p$ is the $p \times p$ identity matrix, and $\boldsymbol{\Psi}=(\gamma_1,...,\gamma_K,\boldsymbol{\theta}_1,...,\boldsymbol{\theta}_K,\sigma_1,...,\sigma_K)$ is the vector of the model's parameters, where $\gamma_1,...,\gamma_K$  with $\gamma_k>0$ and $\sum_{k=1}^K\gamma_k=1$, are the mixture proportions.   Besides,  $K$ is assumed to be known and corresponds to the number of clusters in our application, while $\boldsymbol{\mu}(\boldsymbol{t};\boldsymbol{\theta}_k)=(\mu(t_1;\boldsymbol{\theta}_k),...,\mu(t_p;\boldsymbol{\theta}_k)), 1 \le k \le K,$ being
$\mu(t_j;\boldsymbol{\theta}_k), \forall j \in \{1,...,p\},$ combinations of FMM waves defined as in equation  (\ref{ec:1}). 

The log-likelihood of (\ref{eq:mixFmmDensity}) for a sample of size $n$ is given by:
\begin{equation}
	\log L(\boldsymbol{\Psi})=\sum_{i=1}^n \log \left( \sum_{k=1}^K \gamma_k \;  \mathcal{N}(\boldsymbol{x}_i; \boldsymbol{\mu}(\boldsymbol{t};\boldsymbol{\theta}_k), \sigma^2_k \boldsymbol{I}_p) \right) \label{eqLogV}
\end{equation}

\subsubsection{MLE via EM algorithm}
An EM algorithm is designed to find the MLE of the MixFMM$_m$ model following the methodology in \cite{Dem77, Cha19}. As equation (\ref{eqLogV}) cannot be maximized in a closed form, the complete-data log-likelihood is maximized given the observed data, which is defined as follows:
\begin{equation}
	\log L_c(\boldsymbol{\Psi})=\sum_{i=1}^n \sum_{k=1}^K Z_{ik} \log (\gamma_k \;  \mathcal{N}(\boldsymbol{x}_i; \boldsymbol{\mu}(\boldsymbol{t};\boldsymbol{\theta}_k), \sigma^2_k \boldsymbol{I}_p)) 
\end{equation}
where $Z_{ik} \in \{1,...,K\}, \forall i \in \{1,...,n\}$ and $1 \le k \le K,$ is an indicator variable such that $Z_{ik}=1$ if $\boldsymbol{x}_i$ has been generated by the $k$th FMM model. As described below, the EM starts with an initial solution $\boldsymbol{\Psi}^{(0)}$ and alternates iteratively two steps, maximization and expectation, until convergence is attained. The E-Step computes the expectation of the complete-data log-likelihood given the observed data and the current $\boldsymbol{\Psi}$; whereas in the M-Step, $\boldsymbol{\Psi}$ is updated to maximize the expectation of the complete-data log-likelihood.

\textbf{E-Step} \\
The expectation of the complete-data log-likelihood given the observed data and the current parameter vector $\boldsymbol{\Psi}^{(q)}$ is defined as:
\begin{equation} \label{qFun}
Q(\boldsymbol{\Psi};\boldsymbol{\Psi}^{(q)})=E[\log L_c(\boldsymbol{\Psi})|\boldsymbol{x}_1,...,\boldsymbol{x}_n,\boldsymbol{\Psi}^{(q)}]=\sum_{i=1}^n\sum_{k=1}^K \tau_{ik}^{(q)} \log(\gamma_k^{(q)}\mathcal{N}(\boldsymbol{x}_i; \boldsymbol{\mu}(\boldsymbol{t};\boldsymbol{\theta}_k), \sigma^2_k \boldsymbol{I}_p))
\end{equation}
where $\tau_{ik}^{(q)}$ are the posterior probability of curve $i$ being generated by the $k$th FMM model in the iteration $q$, defined as:
\begin{equation}
\tau_{ik}^{(q)}=P(Z_{ik}=1|\boldsymbol{x}_i; \boldsymbol{\Psi}^{(q)})=
\frac{\gamma_k^{(q)}\mathcal{N}(\boldsymbol{x}_i; \boldsymbol{\mu}(\boldsymbol{t};\boldsymbol{\theta}_k), \sigma^2_k \boldsymbol{I}_p)}{\sum_{h=1}^K\gamma_h^{(q)} \mathcal{N}(\boldsymbol{x}_i; \boldsymbol{\mu}(\boldsymbol{t};\boldsymbol{\theta}_h), \sigma^2_h \boldsymbol{I}_p)}, \; \forall i \in \{1,...,n\}, \;  1 \le k \le K
\end{equation}

\textbf{M-Step} \\
The values of the parameter vector are updated maximizing equation (\ref{qFun}):
\begin{equation} \label{mStep}
\boldsymbol{\Psi}^{(q+1)}=\mathrm{argmax}_{\boldsymbol{\Psi}} Q(\boldsymbol{\Psi};\boldsymbol{\Psi}^{(q)})
\end{equation}
The estimators of the FMM parameters that solve equation (\ref{mStep}) are updated as follows:
\begin{equation} \label{mStep_fmm}
\boldsymbol{\theta}_k^{(q+1)}=\mathrm{argmin}_{\boldsymbol{\theta} \in \Theta} ||\boldsymbol{\overline{x}}^{(q)}_k-\boldsymbol{\mu}(\boldsymbol{t};\boldsymbol{\theta})||^2, \quad 1 \le k \le K
\end{equation}
where $\boldsymbol{\overline{x}}^{(q)}_k=\dfrac{\sum_{i=1}^n\tau_{ik}^{(q)}\boldsymbol{x}_i}{\sum_{i=1}^n\tau_{ik}^{(q)}}$ is the mean waveform for cluster $k$ in the iteration $q$. Equation (\ref{mStep_fmm}) is solved using the standard FMM backfitting algorithm \cite{Rue21a}. 

Then, the estimators of the standard deviation terms are obtained as follows:
\begin{equation}
\sigma_k^{(q+1)}=\sqrt{\frac{1}{p\sum_{i=1}^n\tau_{ik}^{(q)}}\sum_{i=1}^n\tau_{ik}^{(q)}||\boldsymbol{x}_i-\boldsymbol{\mu}(\boldsymbol{t};\boldsymbol{\theta}_k^{(q+1)})||^2}, \quad 1 \le k \le K \label{eqSigma}
\end{equation}

Finally, the cluster proportions are updated as follows:
\begin{equation}
\gamma_k^{(q+1)}=\frac{\sum_{i=1}^n\tau_{ik}^{(q)}}{n}, \quad 1 \le k \le K
\end{equation}
The homoscedasticity restriction $\sigma=\sigma_1=...=\sigma_K$, often assumed in SS problems, reduces 
equation (\ref{eqSigma}) to a single one.
\\

Our proposal for the initial values is: $\gamma_k^{(0)}=1/k, 1 \le k \le K$, and  $\boldsymbol{\theta}^{(0)}_1,...,\boldsymbol{\theta}^{(0)}_K$ are the FMM parameters obtained by fitting FMM$_m$ models  to the mean waveforms of a random cluster assignation.

Success in terms of convergence to the MLE is not initially guaranteed, although the solution converges to a local minimum. We propose to initialize several times randomly in order to avoid local minimums. We have included numeric studies in the paper that show how the proposal reaches a good solution in practice.

The predicted label for each curve can be obtained by assigning the curve to the cluster with the highest estimated posterior probability. It is denoted as $I_i$, and defined as follows $I_i=\mathrm{argmax}_{1 \leq k \leq K} \; \tau_{ik}^{(q)}, \forall i \in \{1,...,n\}$.

\subsection{Selection of the number of clusters}
We propose to determine the number of clusters using a slope heuristic approach, similar to those proposed in  \cite{Bir07, Bou15, Gar15}. Specifically, the log-likelihood $\log L(\boldsymbol{\widehat{\Psi}})$ values for different numbers of clusters starting from one are calculated and plotted consecutively. An exponential trend is expected at the beginning. The minimum value for which the exponential is reduced to a linear trend is considered the optimal number of clusters.  In case of doubt, we suggest breaking ties using additional alternative indexes.

\subsection{Validity metrics}
The procedure for evaluating the goodness of a clustering algorithm is important in order to avoid finding random patterns in data, as well as to compare two clustering partitions. Several works in the literature deal with this question, such as \cite{Ven10, clusterCrit}. Clustering indexes can be classified as external or internal. The former uses externally known information, mainly the class labels; whereas the latter, suitable when the true class labels are unknown, only uses the internal information of the clustering process to evaluate its performance.  In this paper, we consider the accuracy as external index calculated by crossing the predicted and real labels in a confusion matrix. As internal indexes, we consider the Ball-Hall index ($\mathcal{BH}$) and the Davies-Bouldin index ($\mathcal{DB}$), both often considered in SS applications, as well as the average silhouette width ($\mathcal{ASW}$); this being the most simple, interpretable and widely used internal index \cite{Vee20,Akh20}. The definitions of the four indexes are included below to facilitate the interpretation of the results.

Let $\boldsymbol{D}$ be the $r \times r$ confusion matrix obtained from crossing predicted and real classes, in such a way that the diagonal frequencies sum is maximum. The accuracy is defined as:
\begin{equation}
\mathrm{Accuracy}=\frac{\sum_{k=1}^r d_{kk}}{n}
\end{equation}

Now, let $n_k$, $\boldsymbol{c}^k=\dfrac{1}{n_k} \sum_{i/I_i=k} \boldsymbol{x}_i,$ and $\delta^k=\dfrac{1}{n_k} \sum_{i/I_i=k} ||\boldsymbol{x}_i-\boldsymbol{c}^k||^2$ denote, respectively the number of observations, barycenter and dispersion of a cluster $k$.

The $\mathcal{BH}$ is the mean of the clusters' dispersion:
\begin{equation}
\mathcal{BH}=\frac{1}{K} \sum_{k=1}^K \delta^k,
\end{equation}

the $\mathcal{DB}$ is the mean of each cluster's maximum ratio between the sum of the cluster dispersions and the distances between their barycenters:
\begin{equation}
\mathcal{DB}=\frac{1}{K} \sum_{k=1}^K\max_{k \neq k'}\left( \frac{\delta^k+\delta^{k'}}{||\boldsymbol{c}^k-\boldsymbol{c}^{k'}||} \right),
\end{equation}
and the $\mathcal{ASW}$ is defined as:
\begin{equation}
\mathcal{ASW}=\frac{1}{K} \sum_{k=1}^K \frac{1}{n_k} \sum_{i/I_i=k} s_i,
\end{equation}
where $s_i=\frac{b_i-a_i}{\max(b_i,a_i)}$ is the silhouette of  observation $i$, where $a_i=\frac{1}{n_k-1} \sum_{i,i'/I_i=k, I_{i'}=k} ||\boldsymbol{x}_i-\boldsymbol{x}_{i'}||, \forall i \in \{1,...,n\},$ and $b_i=\min_{k' \neq I_i} \left( \sum_{i'/I_{i'}=k'} ||\boldsymbol{x}_i-\boldsymbol{x}_{i'}|| \right), \forall i \in \{1,...,n\}$.

The better the cluster partitions, the lower the expected values of $\mathcal{BH}$ and $\mathcal{DB}$ and the higher the expected values of the accuracy and $\mathcal{ASW}$.

\subsection{Programming languages}
Data from the Allen Cell Types Database was obtained using Python and the Allen SDK \cite{AllenSDK}. The rest of the experimentation was developed with R, including spike detection, spike segmentation, and functional clustering. The MixFMM approach has been implemented using the functions from the FMM package \cite{Fer22}. The PCA and KM implementations used are from the package stats \citep{stats}, whereas the GM are from mclust \cite{mclust}. Moreover, the package clusterCrit \cite{clusterCrit} was used to calculate the clustering internal indexes, while ggplot2 \cite{ggplot2} and plotly \cite{plotly} were used to create the visualizations of the manuscript.

The developed algorithm for the estimation of the MixFMM$_m$ models has been implemented in R and is readily available in the Supplementary Material. Furthermore, a vignette illustrating its basic use is included.

\section{Spike Sorting Datasets}
\label{sec:ssDatasets}

\begin{figure}[!ht]
\begin{center}
\vspace{-5mm}
\includegraphics[width=0.95\textwidth]{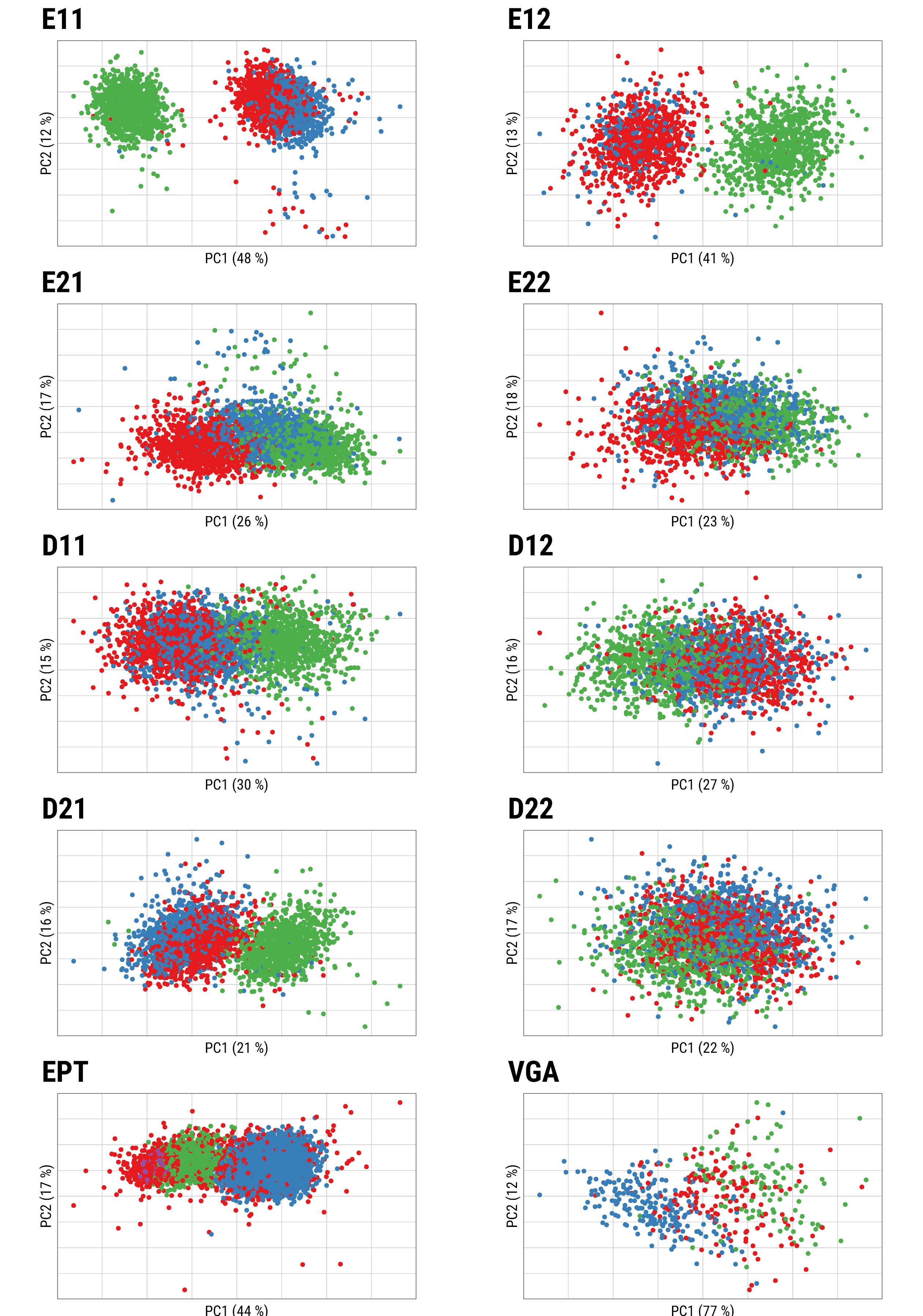}
\end{center}
\vspace{-10mm}
\caption{First two principal components projections of the spikes in the datasets coloured by class. Components' explained variance is shown in the corresponding axis. \label{fig:Figure3}}
\vspace{-5mm}
\end{figure}

\begin{figure}[!ht]
\begin{center}
\makebox[\textwidth]{\includegraphics[width=1.12\textwidth]{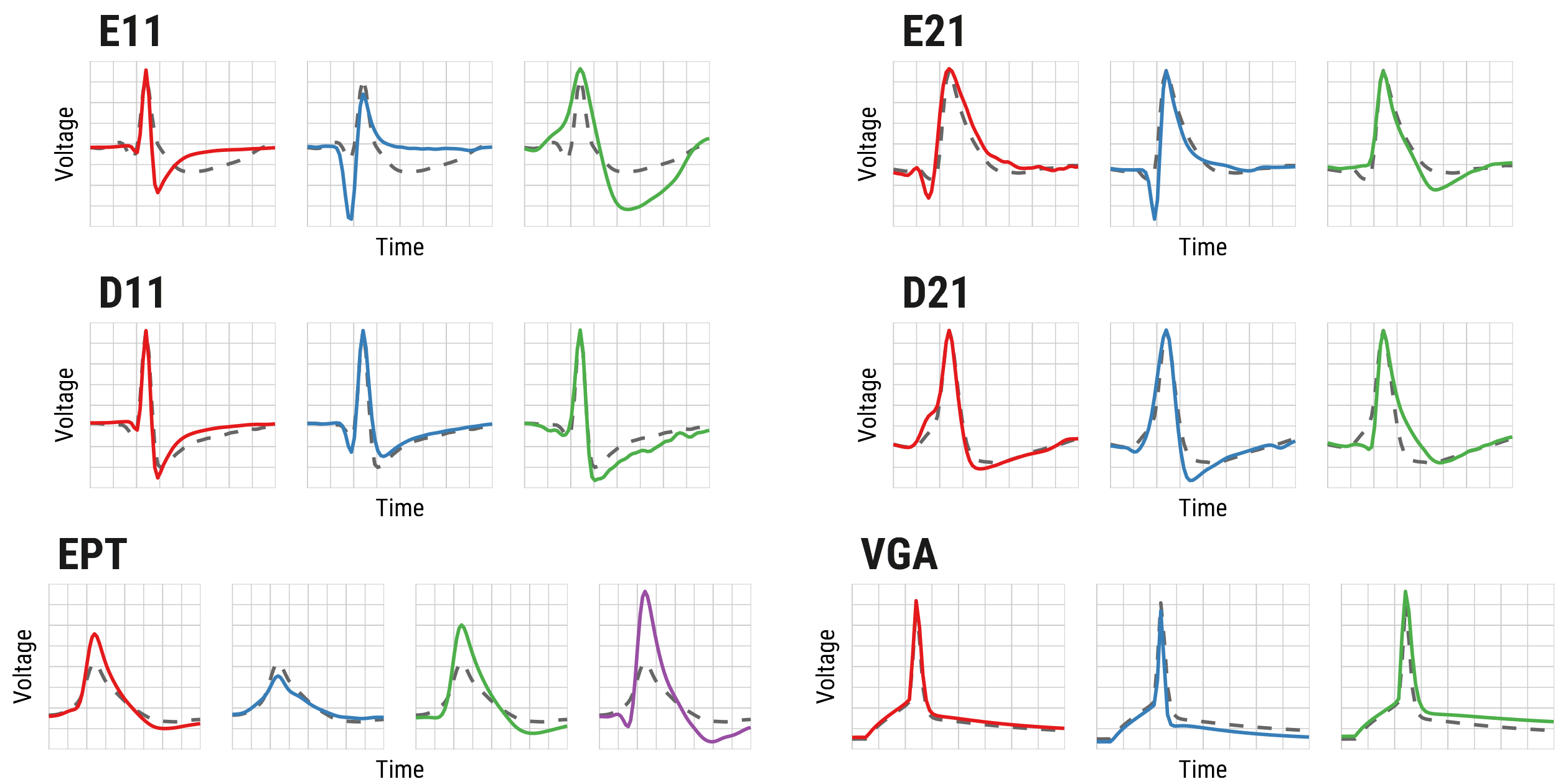}}
\end{center}
\caption{Spike means by class (solid lines) and global spike means (dashed) of datasets.  \label{fig:Figure4}}
\end{figure}

In this work, ten different datasets have been analyzed. They were selected because they encompass a wide variety of spike waveforms -real and simulated-; the real class labels (groundtruth) are known; and they are widely used as benchmarking data in the literature, as in \cite{Sou19, Wan19, Mog19, Vee20}, among others. Table~\ref{table:Datasets} gives information on the number of clusters and signals, and the main references. Furthermore, 2D descriptions are shown in Figure \ref{fig:Figure3}  using the first two principal components; while the mean spike waveform by class (in solid line), along with the global mean (dashed line), are displayed in Figure \ref{fig:Figure4}.

\begin{table}[!ht]
\caption{Summary of the used datasets: name, spike waveform similarity, total number of spikes, number of spikes per class, and origin of the dataset. \label{table:Datasets}}
\begin{center}
\makebox[\textwidth]{%
\begin{tabular}{ c c c c}
\toprule
   & \textbf{Dataset} & \textbf{Nº Spikes} & \textbf{Origin} \\ \toprule
   
 \textbf{E11} & Easy1$\_$Noise01 & $\mathrm{n}=3396$ & \cite{simData}\\[-0.5em]
   & Waveform similarity: Low  & $\mathrm{n}_1=1123, \mathrm{n}_2=1061, \mathrm{n}_3=1212$ &  \\
   
 \textbf{E12} & Easy1$\_$Noise02 & $\mathrm{n}=1896$ & \cite{simData}\\[-0.5em]
   & Waveform similarity: Low  & $\mathrm{n}_1=824,\mathrm{n}_2=193,\mathrm{n}_3=879$ &  \\
      
 \textbf{E21} & Easy2$\_$Noise01 & $\mathrm{n}=3389$ & \cite{simData}\\[-0.5em]
   & Waveform similarity: Middle  & $\mathrm{n}_1=1125, \mathrm{n}_2=1092, \mathrm{n}_3=1172$ &  \\
      
 \textbf{E22} & Easy2$\_$Noise02 & $\mathrm{n}=2707$ & \cite{simData}\\[-0.5em]
   & Waveform similarity: Middle  & $\mathrm{n}_1=951,\mathrm{n}_2=898,\mathrm{n}_3=858$ &  \\
      
 \textbf{D11} & Difficult1$\_$Noise01 & $\mathrm{n}=3365$ & \cite{simData}\\[-0.5em]
   & Waveform similarity: High  & $\mathrm{n}_1=1136, \mathrm{n}_2=1126, \mathrm{n}_3=1103$ &  \\
      
 \textbf{D12} & Difficult1$\_$Noise02 & $\mathrm{n}=2449$ & \cite{simData}\\[-0.5em]
   & Waveform similarity: High  & $\mathrm{n}_1=799, \mathrm{n}_2=792, \mathrm{n}_3=858$ &  \\
      
 \textbf{D21} & Difficult2$\_$Noise01 & $\mathrm{n}=3384$ & \cite{simData}\\[-0.5em]
   & Waveform similarity: High  & $\mathrm{n}_1=1164, \mathrm{n}_2=1119, \mathrm{n}_3=1101$ &  \\
      
 \textbf{D22} & Difficult2$\_$Noise02 & $\mathrm{n}=2634$ & \cite{simData}\\[-0.5em]
   & Waveform similarity: High  & $\mathrm{n}_1=856, \mathrm{n}_2=915, \mathrm{n}_3=863$ &  \\ \midrule
   
    \textbf{EPT} & Epileptic patient temporal lobe & $\mathrm{n}=9195$ &\cite{realData}\\[-0.5em]
   & Waveform similarity: High  &  $\mathrm{n}_1=1050, \mathrm{n}_2=6638, \mathrm{n}_3=1449, \mathrm{n}_4=58$ &  \\
      
 \textbf{VGA} & Visual cortex GABAergic & $\mathrm{n}=507$ & \cite{actData}\\[-0.5em]
   & Waveform similarity: High  & $\mathrm{n}_1=163, \mathrm{n}_2=221, \mathrm{n}_3=123$ & \\ 
   \bottomrule
\end{tabular}
}
\end{center}
\end{table}

More specifically, the eight first datasets are simulated extracellular recordings first presented in \cite{Qui04}. The spikes fired by three different neurons are superimposed on background noise generated by small-amplitude spikes, mimicking the firing of distant neurons. In each dataset pair (E11 and E12, E21 and E22, etc.), the same three neurons are fired with mid and high background noise levels. The lower signal-to-noise ratio generates confusion between classes, as can be seen in Figure \ref{fig:Figure3}. Furthermore, these datasets are ordered in terms of increasing difficulty of correct clustering due to the similarity of the waveforms, as seen in Figure \ref{fig:Figure4}. The data   have been preprocessed as in the cited works: spikes are detected with a threshold, $k=\mathrm{Med}(|\boldsymbol{z}|/0.6745)$ where $\boldsymbol{z}$ is the filtered signal, and the whole voltage traces have been fragmented into signals of 64 samples, each containing a single spike with its maximum aligned with the point 20. The noise artifacts erroneously detected as spikes, less than $4\%$ of the signals, have initially been discarded for the rest of the study.

Finally, the datasets EPT and VGA correspond to real recordings. EPT is a collection of extracellular spikes from four neurons of the temporal lobe of an epileptic patient, openly available in \cite{realData}. In this case, no preprocessing has been done, as the spikes are already cut into signals, with 64 samples and maxima aligned with point 20. In addition, the VGA has been created with data from \cite{actData}, a repository of intracellular electrophysiological neuronal recordings. We selected signals from three GABAergic neuron types (Pvalb, Htr3a and Vip positive) of the mouse visual cortex; in particular, those generated by the short square stimulus with the lowest stimulus amplitude that elicited a single spike for each cell in the database. To facilitate the comparison, the signals have been down-sampled to have 64 samples, as in other datasets, aligning with the maximum at data point 20. Even though the mean spike waveforms are highly similar, as can be seen in Figure \ref{fig:Figure4}, the distinction of GABAergic neuron spikes is crucial as they radically differ in terms of physiological function \cite{Gou20}.

\section{Numerical Studies}
\label{sec:NumStudies}

\begin{figure}[!ht]
\begin{center}
\vspace{-5mm}
\includegraphics[width=\textwidth]{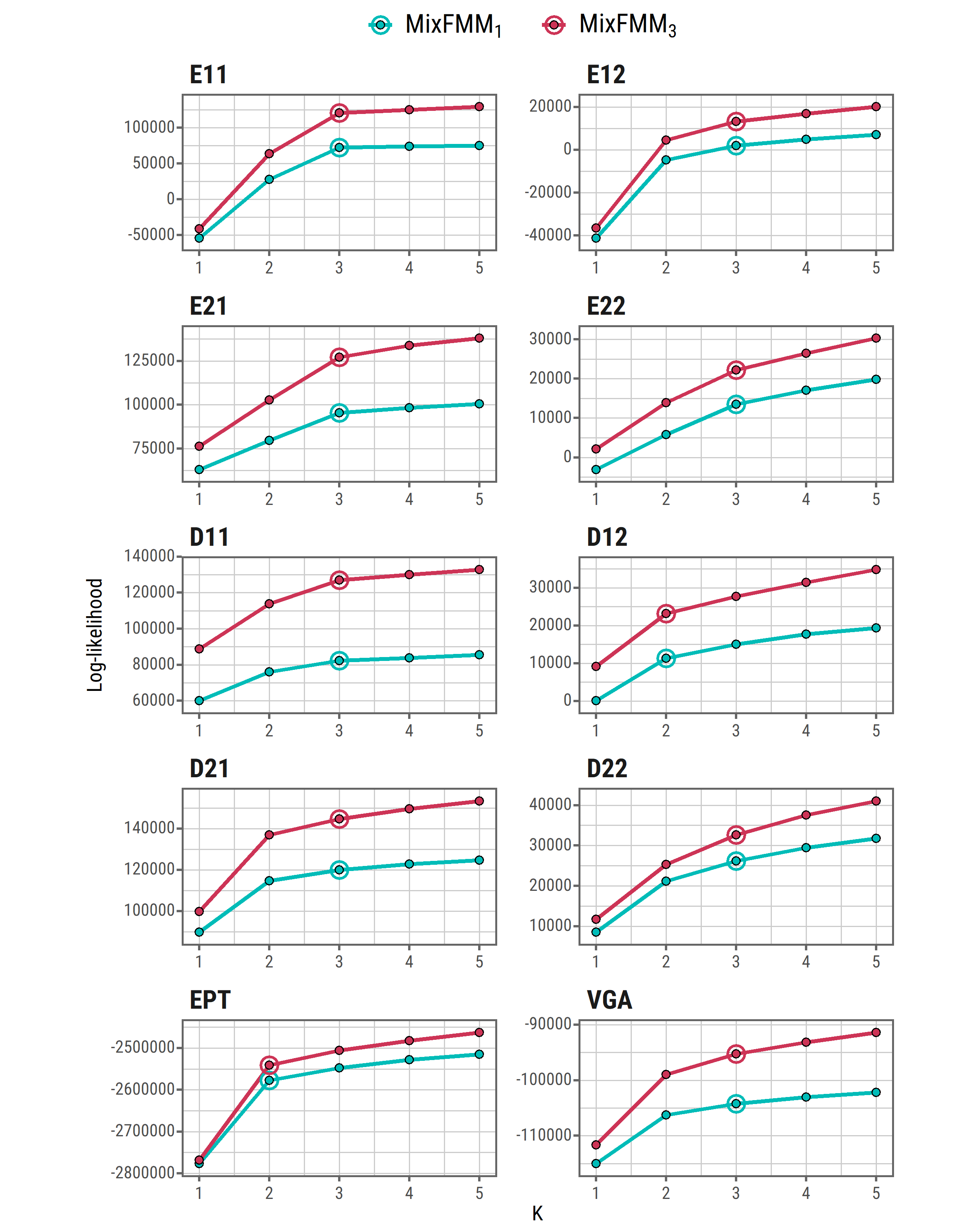}
\end{center}
\vspace{-5mm}
\caption{Log-likelihood of the MixFMM$_1$ and MixFMM$_3$ models for $K=\{2,3,4,5\}$ in each of the datasets. The selected number of clusters is highlighted with a circle. \label{fig:Figure5}}
\end{figure}

In this Section, the MixFMM approach and the PCA (4 components) plus either KM (method PCA+KM) or GM (method PCA+GM), have been used to find clusters in the ten datasets described above.
Specifically, the MixFMM$_1$ and MixFMM$_3$ homoscedastic models have been considered. A preliminary analysis with heteroscedastic models resulted in quite similar variance estimators across the datasets, as well as less accurate and more unstable solutions than those of the homoscedastic models.
 PCA+KM and PCA+GM  have been considered as they are the most widely used clustering algorithms in SS, with better results in practice than wavelets, among other approaches, see \cite{Qui04, Eka14, Vee20}.  The number of principal components in PCA ranges from two to four in the different SS applications. Here, only the results using four components are shown, as this choice gives better results in nine out of the ten datasets. For each of the clustering procedures and datasets, the best model from multiple initializations has been selected. Only ten initial values are needed for the MixFMM and PCA+KM approaches to obtain stable solutions; while the PCA+GM  needed fifty initializations. Furthermore, the optimal number of clusters has been  estimated with the most widely used method for each clustering procedure. A slope BIC heuristic for PCA+KM, and a combined integrated complete-data likelihood criterion plus BIC  for PCA+GM, as proposed in \cite{Pel00, mclust}, respectively. The log-likelihood based approach, described in Section \ref{sec:Methods}, has been used for the MixFMM$_m$ models. Figure \ref{fig:Figure5} shows the log-likelihoods for different numbers of clusters and the selection for each dataset. \\
 
\begin{table}[!ht]
\vspace{-5mm}
\caption{Estimated number of clusters, and index results for each dataset and method.\label{table:Results}
}
\begin{center}
\renewcommand{\arraystretch}{0.65}
\makebox[\textwidth]{%
\begin{tabular}{ccccccccccc} 
\toprule

& \multicolumn{10}{c}{\textbf{Number of clusters}} \\
\cmidrule{2-11}

 & \textbf{E11} & \textbf{E12} & \textbf{E21} & \textbf{E22} & \textbf{D11} & \textbf{D12} & \textbf{D21} & \textbf{D22} & \textbf{EPT} & \textbf{VGA} \\ \midrule

\textbf{PCA+KM} & \textbf{3} & 2 & \textbf{3} & \textbf{3} & \textbf{3} & \textbf{3} & \textbf{3} & \textbf{3} & 3 & \textbf{3} \\

\textbf{PCA+GM} & 4 & 4 & 4 & 4 & 4 & 2 & \textbf{3} & \textbf{3} & \textbf{4} & \textbf{3} \\

\textbf{MixFMM$_1$} & \textbf{3} & \textbf{3} & \textbf{3} & \textbf{3} & \textbf{3} & 2 & \textbf{3} & \textbf{3} & 2 & \textbf{3} \\ 

\textbf{MixFMM$_3$} & \textbf{3} & \textbf{3} & \textbf{3} & \textbf{3} & \textbf{3} & 2 & \textbf{3} & \textbf{3} & 2 & \textbf{3} \\ \midrule

& \multicolumn{10}{c}{\textbf{Accuracy}} \\
\cmidrule{2-11}

 & \textbf{E11} & \textbf{E12} & \textbf{E21} & \textbf{E22} & \textbf{D11} & \textbf{D12} & \textbf{D21} & \textbf{D22} & \textbf{EPT} & \textbf{VGA} \\ \midrule

\textbf{PCA+KM} & 0.992  & 0.888 & 0.875 & 0.563 &
0.665 & 0.524 & 0.720 & 0.540 &
0.589 & 0.617 \\

\textbf{PCA+GM} & 0.960 & 0.958 & 0.851 & 0.347 &
0.905 & 0.510 & 0.640 & \textbf{0.613} &
0.680 & 0.710\\

\textbf{MixFMM$_1$} & 0.992 & 0.966 & 0.936 & 0.700 &
0.895 & 0.530 & 0.705 & 0.575 & 
\textbf{0.874} & \textbf{0.755} \\

\textbf{MixFMM$_3$} & \textbf{0.994} & \textbf{0.976} & \textbf{0.966} & \textbf{0.761} &
\textbf{0.941}& \textbf{0.541} & \textbf{0.743}& 0.606 &
\textbf{0.874} & 0.746 \\ \midrule

& \multicolumn{10}{c}{\textbf{Ball-Hall Index}} \\
\cmidrule{2-11}

 & \textbf{E11} & \textbf{E12} & \textbf{E21} & \textbf{E22} & \textbf{D11} & \textbf{D12} & \textbf{D21} & \textbf{D22} & \textbf{EPT} & \textbf{VGA} \\ \midrule

\textbf{PCA+KM} & 1.153 & 3.344 & 1.162 & 2.852 &
1.182 & \textbf{2.492} & 0.967 & 2.471 &
\textbf{18901} & 1400 \\

\textbf{PCA+GM} & 3.276 & 3.000 & 1.984 & 3.855 &
1.841 & 2.681 & 2.156 & 2.783 &
52804 & 1541 \\

\textbf{MixFMM$_1$} & 1.154 & 2.971 & 1.102 & 2.807 &
1.088 & 2.697 & 0.967 & 2.542 &
21850 & 1230 \\

\textbf{MixFMM$_3$} & \textbf{1.150} & \textbf{2.887} & \textbf{1.085} & \textbf{2.770} &
\textbf{1.074}& 2.688 & \textbf{0.944} & \textbf{2.460} &
21734 & \textbf{1228}\\ \midrule

& \multicolumn{10}{c}{\textbf{Davis-Bouldin Index}} \\
\cmidrule{2-11}

 & \textbf{E11} & \textbf{E12} & \textbf{E21} & \textbf{E22} & \textbf{D11} & \textbf{D12} & \textbf{D21} & \textbf{D22} & \textbf{EPT} & \textbf{VGA} \\ \midrule

\textbf{PCA+KM} & 0.727 & \textbf{0.936} & 1.455 & 2.159 &
2.531 & 2.316 & 2.144 & \textbf{2.103} &
1.601 & 1.213 \\

\textbf{PCA+GM} & 2.411 & 1.140 & 2.421 & \textbf{1.735} &
3.149 & 2.118 & 3.291 & 2.883 &
2.355 & 1.580 \\

\textbf{MixFMM$_1$} & 0.728 & 1.217 & 1.357 & 2.070 &
1.785 & 2.040 & 2.091 & 2.241 &
0.744 & \textbf{1.144}\\

\textbf{MixFMM$_3$} & \textbf{0.726} & 1.159 & \textbf{1.341} & 1.993 &
\textbf{1.717}& \textbf{2.024}& \textbf{2.070} & 2.112 & 
\textbf{0.742} & \textbf{1.144}\\  \midrule

& \multicolumn{10}{c}{\textbf{Average Silhouette Width}} \\
\cmidrule{2-11}

 & \textbf{E11} & \textbf{E12} & \textbf{E21} & \textbf{E22} & \textbf{D11} & \textbf{D12} & \textbf{D21} & \textbf{D22} & \textbf{EPT} & \textbf{VGA} \\ \midrule

\textbf{PCA+KM} & \textbf{0.552} & \textbf{0.439} & 0.250 & 0.117 &
0.126 & 0.120 & 0.173 & 0.128 &
0.262 & 0.280 \\

\textbf{PCA+GM} & 0.348 & 0.356 & 0.160 & \textbf{0.172} &
0.108 & 0.151 & 0.147 & 0.098 &
0.112 & 0.253 \\

\textbf{MixFMM$_1$} & 0.551 & 0.348 & 0.282 & 0.124 &
0.193 & 0.162 & 0.173 & 0.121 &
0.512 & \textbf{0.326} \\

\textbf{MixFMM$_3$} & \textbf{0.552} & 0.365 & \textbf{0.293} & 0.134 &
\textbf{0.205}& \textbf{0.164}& \textbf{0.187} & \textbf{0.133} &
\textbf{0.514} & 0.325 \\ 
 \bottomrule
\end{tabular}
}
\vspace{-5mm}
\end{center}
\end{table}

Table \ref{table:Results} summarizes the main results across different approaches, specifically the optimal number of clusters and the values of the validation indexes. The optimal number of clusters with the MixFMM corresponds to the real clusters in  eight out of ten cases, which is comparable to or better than those of the other approaches. Note that the incorrect cases are characterized by a high level of noise. In terms of accuracy, which is likely the most relevant index, the MixFMM approach attains the best results in nine out of the ten datasets, achieving significant improvements over the PCA-based methods in some cases. The exception is dataset D22, in which PCA+GM is marginally better. Interestingly, PCA+KM achieves better results than PCA+GM, except in datasets where the average curves of the different classes are quite similar.
Furthermore, the internal indexes confirm the better performance of the MixFMM approach against the PCA-based methods traditionally used in SS.  More specifically, the low values of $\mathcal{BH}$ and $\mathcal{DB}$ for the MixFMM approaches indicate, on the one hand, that the clusters have a high cohesion, low dispersion and well-separated barycenters. On the other hand, the MixFMM clusters are composed of similar spikes distinguishable from others in near clusters by the high values of the $\mathcal{ASW}$. 

While most of the results highlight the MixFMM$_3$ as the best all-around model, the MixFMM$_1$ should be considered a suitable alternative in specific cases, particularly in data with a high noise level, as the outstanding results in the real datasets demonstrate. 

Comparing the results in Table \ref{table:Results} with results given in \cite{Vee20} in the common datasets, GM attains significantly better results in this work, probably due to R's mclust being more optimized, flexible, and updated (mclust last version: December 2021; Matlab's gmdistribution last version: before 2018). \\

\begin{figure}[!ht]
\begin{center}
\includegraphics[width=1\textwidth]{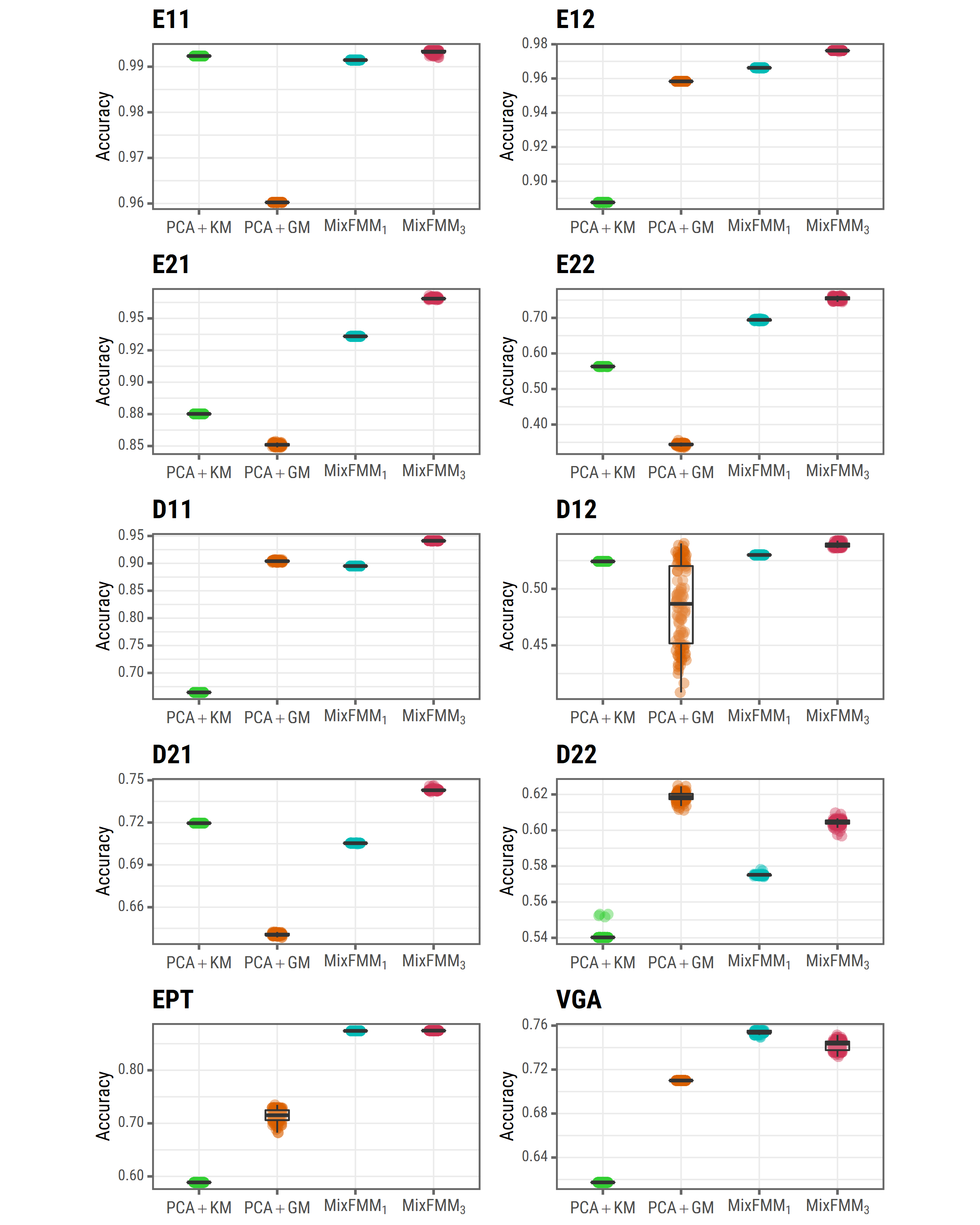}
\end{center}
\vspace{-2mm}
\caption{Boxplots of the accuracy across a hundred repetitions for each dataset and method. \label{fig:Figure6}
}
\end{figure}

Finally, in order to validate the robustness of the different approaches, the distribution of the accuracy has been studied across a hundred repetitions for the methods with ten initializations in each dataset. Figure \ref{fig:Figure6} shows the associated boxplots, which illustrate that the MixFMM approach achieve an accurate solution independently of the initialization in all the scenarios analyzed, while the PCA-based methods fail, especially when spike waveform differences arise mostly from shape and not amplitude.

\section{Discussion}
\label{sec:Discussion}
We present here a novel model-based functional data clustering method and prove its potential to solve the SS problem.  In particular, we show that the MixFMM approach achieves better results than PCA-based methodologies often used in SS applications, achieving significant improvements in some scenarios, including very noisy cases. The MixFMM approach is based on using combinations of FMM waves to describe  the signals, which  allow a wide variety of waveforms to be described in a very simple and interpretable way. In particular, different FMM parameters account for variations in phase, amplitude, and shape. \\

The  results of the MixFMM clustering entails interesting contributions in neuroscience. In particular, neuronal signals from the temporal lobe, such as EPT, are related to specific visual stimuli \cite{Rey15}. The function in the nervous system of the different kinds of GABAergic neurons is radically different, so the distinction of their signals is crucial. In VGA, signals from three types of neurons are differentiated: Pvalb-positive neurons, key in learning and memory processes, as detailed in \citep{Don13}; Vip-positive neurons, related to circadian rhythms \citep{Maz20}; and Htr3a-positive neurons, related to bladder dysfunction \cite{Rit17}. In addition, the methodology  can be adapted to cluster  multichannel spike recordings \cite{Yge18}, combining the information of FMM models for each channel, or with multivariate FMM models similar to those introduced in \cite{Rue22b} to analyze ECG multilead signals. \\
 
Beyond its contribution to neuroscience, the MixFMM is presented as a general approach in functional clustering when oscillatory or quasi-oscillatory signals are the target. In a field in which the literature is ever-growing, and many excessively specific and complex proposals are arising, the need for flexible  and robust proposals is vital. The MixFMM assembles all of these characteristics with a simple but rich parametric formulation that allows differences in phase and/or amplitude to be taken into account. Oscillatory signals appear in astronomy, economy, spectrometry, environmental science, medical imaging, or electrocardiography, among other fields.  In addition, from a theoretical perspective, many extensions of the basic mixture framework could be explored, such as a regularized estimation algorithm or the inclusion of mixed-effects \cite{Cha19}. \\

Finally, one limitation of the proposed methodology is its computational time. It could be significantly reduced by implementing the parameter search in a more computationally efficient programming language, such as C, and using GPU-based computing.

\bibliographystyle{plain}

\end{document}